\documentclass[aps,prd,12pt,superscriptaddress,
amsfonts,amssymb,amsmath,byrevtex,showpacs,nofootinbib]{revtex4-1}
\usepackage{graphicx}
\usepackage{amssymb,amsmath,amsfonts,palatino,amsthm}
\usepackage{amssymb}
\usepackage{epstopdf}
\usepackage{color}

\begin{document}

\title{Big Bang as a critical point\footnote{Essay written for the Gravity Research 
Foundation 2014 Awards for Essays on Gravitation. \\ Submitted on 31 March 2014.}}

\author{Jakub Mielczarek \\
Institute of Physics, Jagiellonian University, Reymonta 4, 30-059 Cracow, Poland \\
Department of Fundamental Research, National Centre for
Nuclear Research,\\ Ho{\.z}a 69, 00-681 Warsaw, Poland}

\email{jakub.mielczarek@uj.edu.pl}

\begin{abstract}

This essay addresses the issue of gravitational phase transitions in the early universe. 
We suggest that a second order phase transition observed in the Causal Dynamical Triangulations 
approach to quantum gravity may have a cosmological relevance. The phase transition 
interpolates between a non-geometric crumpled phase of gravity, and an extended phase  
with classical properties. Transitions of this kind have been postulated earlier in the context 
of geometrogenesis in Quantum Graphity. We show that critical behavior may also be associated 
with a signature change event in Loop Quantum Cosmology. In both cases, classical 
spacetime originates at the critical point associated with a second order phase transition.  
 
\end{abstract}

\maketitle 

\section{Introduction}

Accumulating results of theoretical investigations indicate that the gravitational field  
exists in different phases. First indications supporting such an idea came from 
considerations of three-dimensional Euclidean quantum gravity \cite{Ambjorn:1991rx}. 
By means of Monte Carlo simulations it was possible to explore the configuration 
of the gravitational field under various conditions. For four-dimensional Euclidean
gravity, gravity exhibits two phases: the \emph{crumpled} phase and the 
\emph{branched polymer} phase \cite{Ambjorn:1991pq}.  This result has since been  
generalized to the case of four-dimensional gravity with an imposed causality condition, 
formulation known as Causal Dynamical Triangulations (CDT). The causality condition 
turned out to be essential for the correct phase structure of gravity, leading to emergence
of the four-dimensional spacetime \cite{Ambjorn:2004qm}. Generation of such extended 
phase in the Euclidean approach introducing a non-trivial path integral measure 
remains an interesting possibility \cite{Laiho:2011ya}. Furthermore, in similarity to a phase 
structure of the Lifshitz scalar \cite{Hornreich:1975zz}, the critical surface of CDT has been 
divided into three regions separated by the first and second order transition lines \cite{Ambjorn:2012ij}. 
Interestingly, a theory describing gravity at a triple point (Lifshitz point) of the phase diagram 
has been constructed, and shown to be power-counting renormalizable \cite{Horava:2009uw}. 
Further evidence for the non-trivial phase structure of gravity comes from Quantum 
Graphity \cite{Konopka:2006hu}. This approach utilizes the idea of \emph{geometrogenesis} 
- a transition between geometric and non-geometric phases of gravity. 

The basic question one can ask, assuming the existence of the different phases of 
gravity, is where the other phases can be found? A natural place to search for 
them are high curvature regions such as interiors of the black holes and the 
early universe. Because of a horizon, a possibility of empirical verification of a
phase change inside of black holes is out of reach. More promising is a search 
for signatures of the gravitational phase transitions which took place in the early 
universe. 

So far, there has been very little attention devoted to this issue in the literature.  
Most studies of the phase transitions in the early universe were dedicated 
to the matter sector, rather than gravity \cite{Kibble:1976sj,Zurek:1985qw}.  
Among the few studies  on the gravitational phase transitions in the early 
universe,  the work of Refs.  \cite{Magueijo:2006fu,Dreyer:2013vka} is 
especially noteworthy.  In Ref. \cite{Magueijo:2006fu} a specific model of 
geometrogenesis, through a second order phase transition, has been proposed. 
It was shown that by assuming the holographic principle to be fulfilled in the 
high temperature phase, it is possible to generate a power spectrum of primordial 
perturbations that is in agreement with observations. In Ref. \cite{Dreyer:2013vka} 
the cosmological relevance of second order phase transitions is discussed. 
Arguments supporting generation of ``inflationary'' power spectrum from critical 
behavior of the gravitational field have been presented. 

In what follows we attract attention to the fact that a second order gravitational 
phase transition has recently been observed within Causal Dynamical Triangulations 
\cite{Ambjorn:2011cg}. The transition takes place exactly between the phases 
of the form discussed in Refs.  \cite{Magueijo:2006fu,Konopka:2006hu}. Therefore, 
CDT gives a concrete realization of the scenario of geometrogenesis. We 
also show that gravitational phase transition may be associated with the deformation 
of general covariance, recently observed in the context of Loop Quantum Cosmology 
(LQC). In both cases, the phase transition is of second order, suggesting a 
critical nature of the emergence of classical spacetime in the early universe.  

\section{Causal Dynamical Triangulations} 
 
Analysis performed within four-dimensional CDT with a positive cosmological constant 
indicates the presence of three different phases of the gravitational field, called A, B and C 
\cite{Ambjorn:2012ij}. The phases are separated by the first (A-C) and second order (B-C) 
transition lines presumably intersecting at the triple point. The order of the A-B phase transition 
has not been determined so far. 

At large scales, phase C forms the four-dimensional de Sitter space \cite{Ambjorn:2007jv}. 
The phase is, however, not fully classical since it exhibits dimensional reduction to two 
dimensions at short scales \cite{Ambjorn:2005db}. This can be shown by investigating properties 
of the spectral dimension, defined via a diffusion process. Nevertheless, the phase C 
can be associated with the ``usual" phase of gravity.  The two remaining phases, are 
fundamentally different from this phase. Phase A is characterized by a vanishing interaction 
between adjacent time slices. Phase B, resembling the \emph{crumpled} phase in Euclidean 
gravity, is characterized by a large (tending to infinity in the $\infty-$volume limit) Haussdorf 
and spectral dimension.  Phase B shares features of the high temperature phase postulated 
in Quantum Graphity. Moreover, this phase is separated with the low energetic phase C 
by the second order phase transition. This is in one-to-one correspondence to the 
Quantum Graphity case.  Based on this observation, we hypothesize the following:  

\emph{Hypothesis 1. In the early universe, there is a second order phase transition from 
the high temperature phase B to the low temperature phase C. The transition is associated 
with a change of the connectivity structure between the elementary chunks of space.} 

The change of connectivity can be inferred from the considerations of the spectral 
dimensions of the phases B and C.  In order to see it explicitly let us consider a 
toy model of the universe composed of the $N$ chunks of space. They will be represented 
by the nodes of a graph. A structure of adjacency is represented by the links.   

\begin{figure}[ht!]
\centering
$\begin{array}{cc}
\text{Geometric phase (low temperature)} \ \ \ \ \ \ &\ \ \ \ \
\text{Non-geometric phase (high temperature)}  \\ 
\includegraphics[width=5cm, angle = 0]{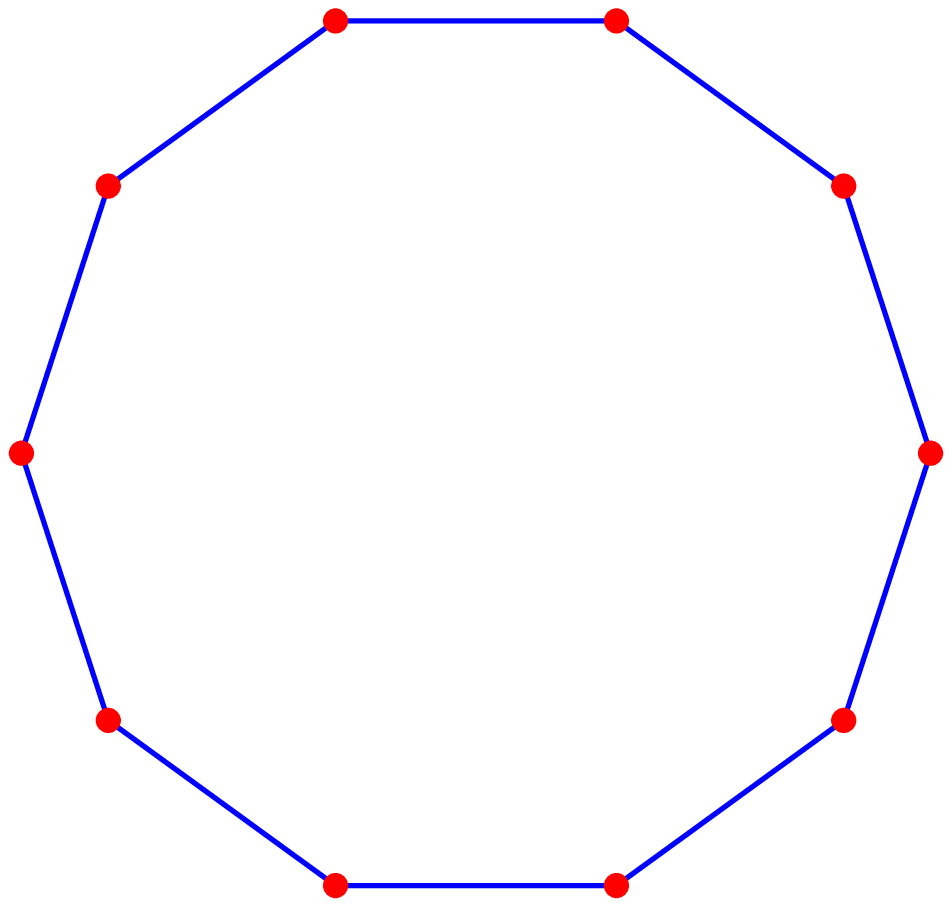}   &
\includegraphics[width=5cm, angle = 0]{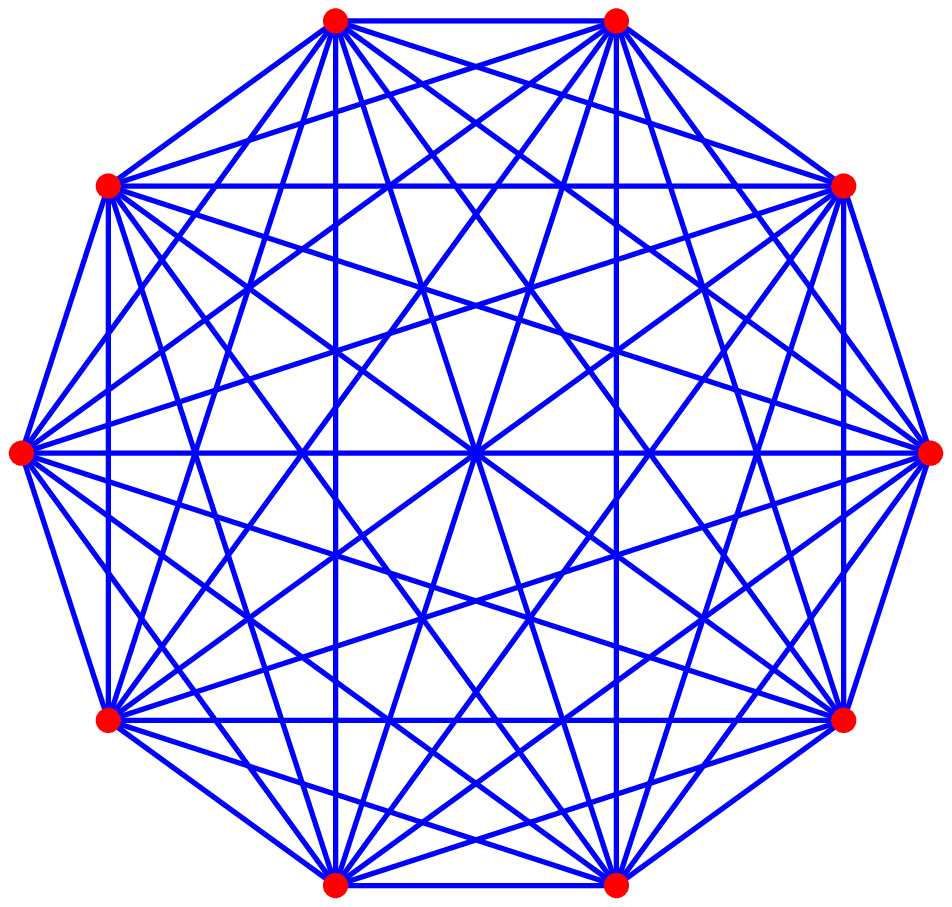} 
\end{array}$
\caption{(Left) Ring graph being a toy model of the low temperature geometric state of gravity. 
(Right)  Complete graph being a model of high temperature non-geometric state of gravity. }
\label{Graphs}
\end{figure}

In phase C, which is a geometric phase, the degree of vertices is low. In our toy model 
it is equal $2$ and the resulting space is represented by the Ring graph (see the left 
panel in Fig. \ref{Graphs}). The spectral dimension of this graph can be found by determining 
spectrum (eigenvalues $\lambda_n$) of the Laplace operator $\Delta \equiv A-D$, where $A$ is 
an adjacency matrix and $D$ is a degree matrix.  By using the expression for the trace 
of the heat kernel one can find that 
\begin{equation}
d_S \equiv   -2 \frac{\partial \log \text{tr} K}{\partial \log \sigma} = 
- 2\sigma \frac{\sum_{i=1}^N \lambda_n e^{\lambda_n\sigma}}{\sum_{i=1}^N e^{\lambda_n\sigma}},
\label{SpectralDim}
\end{equation} 
where $\sigma$ is a diffusion time. 

\begin{figure}[ht!]
\centering
\includegraphics[width=10cm, angle = 0]{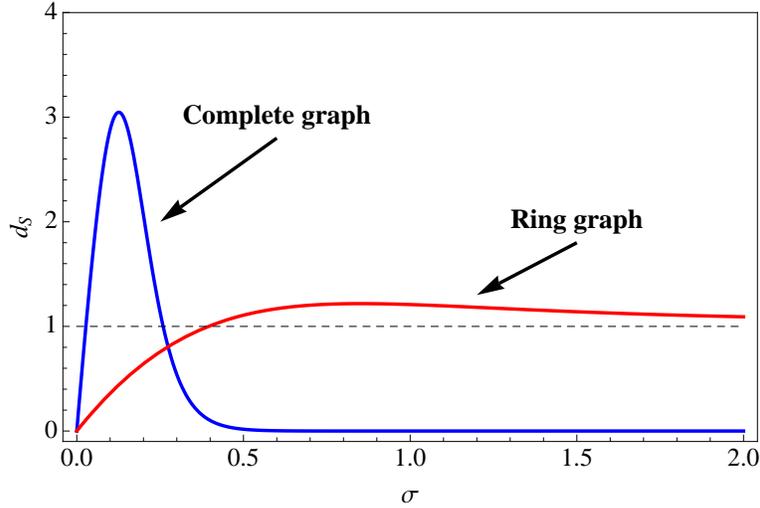}
\caption{Spectral dimensions for the Ring graph (geometric phase) and the Complete graph 
(non-geometric phase) with $N=20$ nodes.}
\label{SpectDim}
\end{figure}

In Fig. \ref{SpectDim} we plot function (\ref{SpectralDim}) for the Ring graph, for which the eigenvalues 
are  $\lambda_n=2 \left( \cos(2\pi n/N)-1\right)$.  At intermediate diffusion times the spectral dimension 
is equal to one, as expected classically. The short time behavior, corresponding to dimensional reduction 
observed in the four-dimensional case, is due to the discrete nature of the network. Furthermore, at large 
diffusion times the spectral dimension is again falling to zero due to the compactness of space (not visible 
in Fig. \ref{SpectDim}).     

Let us now model the high temperature non-geometric phase B by assuming, for computational 
simplicity, a maximal degree of nodes. The resulting Complete graph is shown in  Fig. \ref{Graphs}. 
The assumption of completeness allows one to determine spectrum of the Laplace operator analytically, 
and enables the simplification of Eq. \ref{SpectralDim} to 
\begin{equation}
d_S = \frac{2N \sigma (N-1)}{e^{N\sigma}+(N-1)}.
\end{equation}
We plot this function in Fig. \ref{SpectDim}, comparing it with the low temperature case. As 
we see, the spectral dimension is now peaked at small diffusion times. The maximal value 
of the spectral dimension grows with the number of nodes as $d_{S,\text{max}} \approx 2 W(N/e)$,
where $W(x)$ is the Lamber function.  This behavior is in qualitative agreement with the numerical 
computations performed in four-dimensional CDT \cite{Ambjorn:2012ij, AGPriv}.  
   
\section{Loop Quantum Cosmology}  
 
Recent developments in LQC indicate that the hypersurface deformation algebra 
(HDA) is deformed due to the quantum gravitational effects \cite{Bojowald:2011aa}. 
This means that the general covariance is quantum deformed, but not broken. 

In condensed matter physics, a change of symmetry is often 
associated with the occurrence of a phase transition. By extrapolating this 
observation to the sector of gravitational interactions, we make the following hypothesis: 

\emph{Hypothesis 2. Gravitational phase transitions are associated with 
deformations of the hypersurface deformation algebra.} 
   
In order to support this hypothesis we present an example based on holonomy 
corrections in LQC. In this case, HDA is deformed such that $\{S,S\}=\Omega D$, 
where $S$ are scalar constraints and $D$ is a diffeomorphism constraint, and the 
remaining brackets are unchanged \cite{Cailleteau:2011kr}. The $\Omega=1-2\frac{\rho}{\rho_{c}}$ 
is a deformation factor, $\rho$ denotes energy density of matter and $\rho_c$ is a 
maximal energy density expected to be of the order of the Planck energy density. 

At low energy densities, the classical Lorentzian HDA with $\Omega=1$ is recovered, 
while at $\rho = \rho_{c}$, $\Omega=-1$ corresponding to Euclidean space. 
Approaching the Planck epoch is therefore associated with the signature change 
from Lorentzian spacetime to Euclidean four-dimensional space \cite{Mielczarek:2012pf}. 
Interestingly, at $\rho = \rho_{c}/2$ the HDA reduces to the ultralocal form ($\{S,S\}=0$)
describing a state of \emph{silence} \cite{Mielczarek:2012tn}. This state shares properties 
of phase A in CDT, giving a first indication for the relationship between the phase 
of gravity and deformation of HDA. 
 
Further evidence comes from a simple model of the signature change as a spontaneous 
symmetry breaking (SBB) associated with a second order phase transition. The symmetry
breaking is from $SO(4)$ to $SO(3)$ at the level of an effective homogeneous vector field 
$\phi_{\mu}$. This translates to a symmetry change from $SO(4)$ to $SO(3,1)$, experienced 
by the field living on a geometry described by the metric $g_{\mu\nu} = \delta_{\mu\nu}-2\phi_{\mu} \phi_{\nu}$. 

Let us assume that the free energy for the model with a massless scalar field $v$ is
\begin{eqnarray}
F = \int dV  \underbrace{ \left(  \delta^{\mu\nu}+\frac{2 \phi^{\mu} \phi^{\nu}}{1
-2 |\vec{\phi}|^2}  \right)  }_{g^{\mu\nu}}\partial_{\mu} v \partial_{\nu} v 
+\underbrace{\beta\left[  \left( \frac{\rho}{\rho_{\text{c}}}-1\right)|\vec{\phi}|^2
+\frac{1}{2}|\vec{\phi}|^4 \right]}_{V(\phi^{\mu},\rho)},  \nonumber
\end{eqnarray} 
where $|\vec{\phi}| = \sqrt{\delta^{\mu\nu} \phi_{\mu} \phi_{\nu}}$ and $\beta$ is a constant.
Due to the symmetry breaking kinetic factor, the expression for the free energy is not explicitly 
$SO(4)$ invariant. Treating this term as a perturbation (which is 
valid for sufficiently small values of $v$), the equilibrium is obtained by minimizing 
value of the potential $V(\phi^{\mu},\rho)$. 

\begin{figure}[ht!]
\centering
\includegraphics[width=8cm, angle = 0]{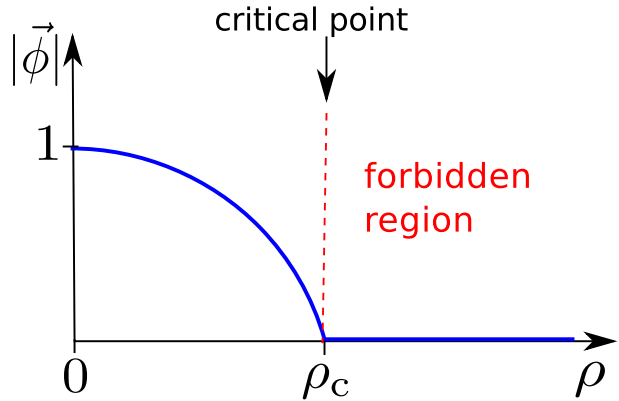}
\caption{Modulus of the filed $\phi^{\mu}$ as a function of the energy density $\rho$.
The region $\rho >\rho_c$ is forbidden within the model.}
\label{SSB}
\end{figure}

At energy densities $\rho > \rho_c$ the vacuum state maintains the $SO(4)$ symmetry, leading 
to $|\vec{\phi}|=0$. The metric elements are therefore $g_{00}=1$ and $ g_{ii}  = 1$, 
representing the four-dimensional Euclidean space. This region is however forbidden 
due to the constraint $\rho \leq \rho_c$ present in LQC. In turn, below the critical energy 
$\rho \leq \rho_{\text{c}}$, the minimum of the potential is located at $|\vec{\phi}|=\sqrt{1-\frac{\rho}{\rho_{\text{c}}}}$
in some spontaneously chosen direction (See Fig. \ref{SSB}).
Without loss of generality, let us assume that the SSB takes place in direction $\phi_0$, for which  
\begin{eqnarray}
g_{00} = 1-2\phi_0\phi_0=1-2\left(1-\frac{\rho}{\rho_{\text{c}}}\right) 
=-1+2\frac{\rho}{\rho_{\text{c}}}=-\Omega, 
\ \ \ g_{ii}  = 1, \nonumber
\end{eqnarray}
leading to the effective speed of light $c_{\text{eff}}^2 =\Omega$. As a consequence, the equation 
of motion for the scalar field $v$ takes the form 
\begin{equation}
g^{\mu\nu}\partial_{\mu}\partial_{\nu}v = -\frac{1}{c_{\text{eff}}^2}\frac{\partial^2}{\partial t^2} v+\Delta v = 0,  
\label{EQMv}
\end{equation}
manifesting $SO(4)$ symmetry at the critical point ($\Omega=-1$), and  $SO(3,1)$ symmetry in the 
low temperature limit ($\Omega=1$). The form of equation (\ref{EQMv}) agrees with the one 
derived from the holonomy deformations of the HDA \cite{Bojowald:2011aa}. 
 
In LQC, energy densities above $\rho_c$ cannot be reached. Therefore, the evolution starts 
at the critical point located at $\rho_c$. An interesting possibility is that the system has been 
maintained at the critical point before the energy density started to drop. This may not 
require a fine-tuning if the dynamics of the system exhibited Self Organized Criticality (SOC)
\cite{Bak:1987xua}, which is observed in various complex systems.  Interestingly, this concept 
has already been applied to quantum gravity, however, in order to describe classical configuration 
of space \cite{Ansari:2004an}. Furthermore, it is possible that the region $\rho >\rho_c$ is 
described by a non-geometric phase of gravity, which properties cannot be captured within the 
presented model. 
 
\section{Conclusions}

We have shown that CDT offers a concrete realization of geometrogenesis, having
a second order gravitational phase transition between the non-geometric and geometric
phase. We have explained how measurements of the spectral dimension are related 
in this case with connectivity of the chunks of space. The critical nature of the 
emergence of classical spacetime in the early universe may give a first possibility 
of testing CDT. However, this would require more detailed investigations of the 
properties of the second order phase transition in CDT.  In LQC, which is an alternative 
to CDT, a gravitational second order phase transition may explain a signature change 
in the Planck epoch. In the presented toy model, the universe originates just at the 
critical point.  Interestingly, the latest results in CDT indicate the existence of a new 
phase between phases B and C \cite{Ambjorn:2014mra}, characterized by a ``bifurcation'' 
of the kinetic term. This behavior resembles the signature change observed in LQC.


\begin{thebibliography}{99}

\bibitem{Ambjorn:1991rx}
  J.~Ambjorn, D.~V.~Boulatov, A.~Krzywicki and S.~Varsted,
  Phys.\ Lett.\ B {\bf 276} (1992) 432.
  
\bibitem{Ambjorn:1991pq}
  J.~Ambjorn and J.~Jurkiewicz,
  Phys.\ Lett.\ B {\bf 278} (1992) 42.
  
\bibitem{Ambjorn:2004qm}
  J.~Ambjorn, J.~Jurkiewicz and R.~Loll,
  Phys.\ Rev.\ Lett.\  {\bf 93} (2004) 131301
  [hep-th/0404156].
  
\bibitem{Laiho:2011ya}
  J.~Laiho and D.~Coumbe,
  Phys.\ Rev.\ Lett.\  {\bf 107} (2011) 161301
  [arXiv:1104.5505 [hep-lat]].
  
\bibitem{Hornreich:1975zz}
  R.~M.~Hornreich, M.~Luban and S.~Shtrikman,
  Phys.\ Rev.\ Lett.\  {\bf 35} (1975) 1678.
  
\bibitem{Ambjorn:2012ij}
  J.~Ambjorn, S.~Jordan, J.~Jurkiewicz and R.~Loll,
  Phys.\ Rev.\ D {\bf 85} (2012) 124044
  [arXiv:1205.1229 [hep-th]].
  
\bibitem{Horava:2009uw}
  P.~Horava,
  Phys.\ Rev.\ D {\bf 79} (2009) 084008
  [arXiv:0901.3775 [hep-th]].
  
\bibitem{Konopka:2006hu}
  T.~Konopka, F.~Markopoulou and L.~Smolin,
  hep-th/0611197.
  
\bibitem{Kibble:1976sj}
  T.~W.~B.~Kibble,
  J.\ Phys.\ A {\bf 9} (1976) 1387.
  
\bibitem{Zurek:1985qw}
  W.~H.~Zurek,
  Nature {\bf 317} (1985) 505.
    
\bibitem{Magueijo:2006fu}
  J.~Magueijo, L.~Smolin and C.~R.~Contaldi,
  Class.\ Quant.\ Grav.\  {\bf 24} (2007) 3691
  [astro-ph/0611695].
  
\bibitem{Dreyer:2013vka}
  O.~Dreyer,
  arXiv:1307.6169 [gr-qc].
  
\bibitem{Ambjorn:2011cg}
  J.~Ambjorn, S.~Jordan, J.~Jurkiewicz and R.~Loll,
  Phys.\ Rev.\ Lett.\  {\bf 107} (2011) 211303
  [arXiv:1108.3932 [hep-th]].
  
\bibitem{Ambjorn:2007jv}
  J.~Ambjorn, A.~G\"orlich, J.~Jurkiewicz and R.~Loll,
  Phys.\ Rev.\ Lett.\  {\bf 100} (2008) 091304
  [arXiv:0712.2485 [hep-th]].
  
\bibitem{Ambjorn:2005db}
  J.~Ambjorn, J.~Jurkiewicz and R.~Loll,
  Phys.\ Rev.\ Lett.\  {\bf 95} (2005) 171301
  [hep-th/0505113].
  
\bibitem{AGPriv}
A.~G\"orlich, private communication.  

\bibitem{Bojowald:2011aa}
  M.~Bojowald and G.~M.~Paily,
  Phys.\ Rev.\ D {\bf 86} (2012) 104018
  [arXiv:1112.1899 [gr-qc]].
  
\bibitem{Cailleteau:2011kr}
  T.~Cailleteau, J.~Mielczarek, A.~Barrau and J.~Grain,
  Class.\ Quant.\ Grav.\  {\bf 29} (2012) 095010
  [arXiv:1111.3535 [gr-qc]].
  
\bibitem{Mielczarek:2012pf}
  J.~Mielczarek,
  arXiv:1207.4657 [gr-qc].
 
\bibitem{Mielczarek:2012tn}
  J.~Mielczarek,
  AIP Conf.\ Proc.\  {\bf 1514} (2012) 81
  [arXiv:1212.3527 [gr-qc]].
  
\bibitem{Bak:1987xua}
  P.~Bak, C.~Tang and K.~Wiesenfeld,
  Phys.\ Rev.\ Lett.\  {\bf 59} (1987) 381.

\bibitem{Ansari:2004an}
  M.~H.~Ansari and L.~Smolin,
  Class.\ Quant.\ Grav.\  {\bf 25} (2008) 095016
  [hep-th/0412307].
  
\bibitem{Ambjorn:2014mra}
  J.~Ambjorn, J.~Gizbert-Studnicki, A.~G\"orlich and J.~Jurkiewicz,
  arXiv:1403.5940 [hep-th].
     
\end{thebibliography}
\end{document}